\documentstyle[11pt,aaspp4]{article}

\tighten
 
\newcommand{\kms}{$\,{\rm km\,s^{\scriptscriptstyle -1}}$}
\newcommand{\gtsim}{\ {\raise-0.5ex\hbox{$\buildrel>\over\sim$}}\ }
\newcommand{\ltsim}{\ {\raise-0.5ex\hbox{$\buildrel<\over\sim$}}\ }

\def\simlt{\lower.5ex\hbox{$\; \buildrel < \over \sim \;$}}
\def\simgt{\lower.5ex\hbox{$\; \buildrel > \over \sim \;$}}
 
\begin{document}

\title{The Faint End of the Luminosity Function of Galaxies in Hickson Groups}

\author{Stephen E. Zepf\altaffilmark{1,2}}
\affil{Dept.\ of Astronomy, Univ.\ of California, Berkeley, CA 
94720, \\
and Dept.\ of Astronomy, Yale University, New Haven, CT 
06520}

\author{Reinaldo R. de Carvalho}
\affil{Observat\'orio Nacional - CNPq - DAF, Brazil}

\centerline{and}

\author{Andr\'e L. B. Ribeiro}
\affil{Divis\~ao de Astrof\'{\i}sica - INPE/MCT, C.P. 515 - 
12201-970 - S. Jos\'e dos Campos, Brazil}

\vskip 24pt

\centerline{To appear in {\it The Astrophysical Journal Letters}}

\vskip 24pt

\altaffiltext{1}{Hubble Fellow}
\altaffiltext{2}{e-mail:zepf@astro.yale.edu}

\begin{abstract}

	We study the luminosity function of galaxies in Hickson groups
using our recent redshift survey of galaxies in and around 17 of these 
groups. We find that the galaxies in these regions have a luminosity function
with $M_{*} = -19.5 + 5 {\rm log} h$, and $\alpha = -1.0$, 
where $M_{*}$ and $\alpha$ are the usual parameters in the standard 
Schechter form of the luminosity function, and the magnitudes are measured 
in the $B$ band. The formal $95\%$ confidence intervals for $M_{*}$ 
and $\alpha$ range from (-19.3,-0.8), to (-19.7,-1.2) and are
highly correlated as is usual for these fits. 
This luminosity function for galaxies in our Hickson
group sample is very similar from that found in large surveys covering
a range of environments. These values are also consistent with our earlier
estimates based on a photometric analysis with statistical background 
correction, and do not support previous suggestions of an underabundance 
of intrinsically faint galaxies in compact groups. We confirm our earlier 
finding that the fainter galaxies are more diffusely distributed within 
individual groups than the brighter ones. This can be interpreted either 
as evidence for mass segregation within the groups or as the result of 
the selection procedure for Hickson groups.
  
\end{abstract}


\section{Introduction} 

	The luminosity function of galaxies is generally
described by a function of the form
$$ n(L) dL \propto e^{-L/L_{*}} (L/L_{*})^{\alpha} dL$$ (Schechter 1976). 
However, there is little consensus on the
value of $\alpha$, the slope of the faint end of the luminosity function.
Published values range from roughly $\alpha \simeq -0.8$
to $\alpha \simeq -2.0$. Moreover, it is not yet clear whether the
slope of the faint end of the luminosity function is dependent on the 
environment of the galaxies sampled.

	


	Possible environmental variations of the faint end slope of the 
galaxy luminosity function are interesting, as they may reflect
astrophysical process during galaxy formation, or subsequent dynamical
evolution. Observations suggest that higher luminosity galaxies are 
more strongly clustered than low luminosity galaxies 
(e.g.\ Lin et al.\ 1996a, Loveday et al.\ 1995). This is at least
qualitatively consistent with the predictions of biased galaxy
formation models in which low density regions preferentially harbor 
low luminosity galaxies (e.g.\ White et al.\ 1987). 
Dynamical effects may also lead
to environmental differences in the luminosity function. For example,
mass segregation will tend to lead towards more massive objects
in denser regions. However, these massive galaxies may also preferentially
disappear through merging as the result of dynamical friction.

	In practice, the slope of the faint end of the luminosity
function is determined in one of two ways. The most straightforward
approach is through large redshift surveys. The largest and most
recent of these is the Las Campanas Redshift Survey, from which
Lin et al.\ (1996b) derive $\alpha = -0.70 \pm 0.05$, for galaxies
in the magnitude range $-22 \gtsim M_{B} + 5 log h \gtsim -16.5$, 
with evidence for a somewhat steeper slope ($\alpha \simeq -1$) when 
the fit is extended to fainter galaxies. The LCRS luminosity function 
is in good agreement with that derived from the earlier Stromlo-APM 
survey, for which Loveday et al.\ (1992) found $\alpha \simeq -1$.
A similarly shallow slope was found in the CfA survey for brighter
magnitudes (Marzke et al.\ 1994a). However, Marzke et al.\ (1994a)
also found a significant steepening of the slope at the faint end 
which is not seen in either of the other surveys.

	An alternative technique is to compare the galaxy counts 
within a cluster to those outside of the cluster, and thereby 
derive statistically a luminosity function for the cluster.
This approach has now been applied to a number of clusters,
with varied results. For example, Gaidos (1997) surveyed 20
Abell clusters and found $\alpha \simeq -1.1$, consistent 
with earlier results of Schechter (1976) and Dressler (1978)
who found fairly flat slopes for the luminosity function of
cluster galaxies. Detailed studies of the Coma cluster 
(Bernstein et al.\ 1994) and the Virgo cluster (Sandage et al.\ 1985) 
reveal somewhat steeper slopes, with $\alpha \simeq -1.4$.
Much steeper slopes ($\alpha \simeq -2$) were found by
De Propis et al.\ (1995) for four low redshift clusters, and
by Driver et al.\ (1994) for several clusters at moderate redshift.
Lopez-Cruz et al.\ (1997) claim a systematic variation from shallow
to steep slopes with decreasing richness of clusters, consistent
with the early work of Oemler (1974), although other studies
have indicated a universal luminosity function for cluster galaxies
(e.g.\ Lugger 1986, Colless 1988).

	One effective way to test for environmental effects on galaxies
is to study galaxies in Hickson groups. These groups were selected
on the basis on their very high surface densities (Hickson 1982).
Subsequent spectroscopic observations have established that most
of the galaxies in individual groups are at similar redshifts,
and that the groups typically have velocity dispersions of 100-350 \kms.
(Hickson et al.\ 1992, see also Ribeiro et al.\ 1997).
The combination of high spatial densities inferred from the
projected galaxy distribution and velocity dispersions similar 
to the internal velocities of galaxies gives short
timescales for dynamical evolution through galaxy merging within
compact groups, as dramatically demonstrated by Barnes (1989). 

	Given the short timescales for dynamical evolution expected
in Hickson groups, it is interesting to compare the luminosity 
function of galaxies in Hickson groups to the general field population.
We previously addressed this problem by counting galaxies in and around
a sample of Hickson groups, and then statistically correcting for
background galaxies (Ribeiro et al.\ 1994). This leads to an estimate
of the faint end of the galaxy luminosity function in much the same
way as the cluster studies described above. With this approach, we were 
able to reach  much fainter magnitudes than considered by Hickson 
in his group selection, thereby avoiding the difficult problem of accurately 
modeling the selection effects in the Hickson sample that led to 
disagreements between earlier studies (Mendes de Oliveira \& Hickson 1991, 
Sulentic \& Rabaca 1994). Reaching fainter magnitudes is also obviously 
valuable for improving the leverage on the determination of the slope 
of the faint end of the luminosity function. 

	The photometric analysis indicated that the faint end of the
luminosity function was well-fit by a Schechter function with
$\alpha = -0.82 \pm 0.09$. The uncertainty reflects the statistical
uncertainty in the number of galaxies detected above the estimated
background. There are potential systematic concerns associated with 
the background corrections. Therefore, one of the motivations for 
our spectroscopic survey of faint galaxies in and around compact groups
(de Carvalho et al.\ 1997) was to eliminate the need for statistical 
background correction by obtaining redshifts for these galaxies.
This paper reports the results of the analysis in $\S 2$, and discusses
the implications of these results in $\S 3$.

\section{Analysis}

	In order to determine the luminosity function of compact groups, 
we utilize redshifts determined in our spectroscopic survey of galaxies 
in and around 17 Hickson groups (de Carvalho et al.\ 1997). $B$ magnitudes 
are obtained from our earlier photometric analysis of galaxies in these 
regions (de Carvalho et al.\ 1994). We then combine the redshifts and 
the photometry to determine the distribution of galaxy luminosities within 
each group. The faint limit of this procedure is taken to be the $B$ magnitude 
at which our redshift survey is $10\%$ incomplete for that group.
In order to determine the luminosity function, we weight the galaxy 
luminosity distribution within each group by the effective volume 
$(v/v_{\rm max})$ of that 
group. For the selection function of the groups, we adopt the form
$P(m) = (1 + 10^{(1.2(m-m_0))})^{-1}$
given by Hickson, Kindl, \& Aumann (1989) for the Hickson group sample.
As described in Ribeiro et al.\ (1994), $m_0 = 13.0$ gives the
best fit to the cumulative distribution of total magnitude of our sample
of groups, which is a subsample of the total Hickson catalog 
(de Carvalho et al.\ 1994). 
The galaxy luminosity function for the sample as a whole is then
determined by a straightforward summation over the 17 groups using
the $v/v_{\rm max}$ weighting for each group. 
Because each group encompasses a wide range of galaxy luminosities,
the shape of the resulting galaxy luminosity function is not sensitive 
to the details of the weighting procedure.
Uncertainties in the effective volume of different groups 
tend to shift the normalization of the luminosity function ($\phi^{*}$), 
but not its shape ($M_{*}$ and $\alpha$).

	The resulting luminosity function for galaxies in our sample
of 17 Hickson groups is given in Figure 1. Also plotted on this figure
is the best fitting Schechter function, which has
$M_{*} = -19.5 + 5 {\rm log} h$ and $\alpha = -1.0$,
as well as $\phi^{*} = 2 \times 10^{-4}$. 
The parameters, $\phi^{*}$, $M_{*}$, and $\alpha$ were determined 
by a non-linear least-squares fit (Jeffreys, Fitzpatrick, \& McArthur 1988). 
The error bars for the individual points were determined by the 
standard deviation (1$\sigma$) of the galaxy counts in each luminosity bin.
As is usual for these fits, the values of $M_{*}$ and $\alpha$ 
are highly correlated, and the formal $95\%$ confidence limits on 
$M_{*}, \alpha$ combinations are (-19.7, -1.2) and (-19.3, -0.8).

	{\it The primary result of this paper is that the luminosity
function of galaxies in our sample of Hickson groups is very similar
to that found in similar surveys of large samples of galaxies covering
a wide range of environments}. This agreement is shown in Figure 2,
where we plot both our luminosity function for galaxies in Hickson
groups and the LCRS luminosity function of Lin et al.\ (1996b).
The $r$ magnitudes of the LCRS have been converted to $B$ magnitudes
using $B - r$ = 1.1 (Lin et al.\ 1996b).
The luminosity functions have been offset arbitrarily in the y-axis
for ease of comparison.

	We also note that the luminosity function derived here from 
our spectroscopic survey is consistent with the one we derived
earlier by comparison galaxy counts inside and outside of the groups
(Ribeiro et al.\ 1994).
This agreement suggests that the statistical background subtraction
adopted in our earlier paper is reliable. It also suggests that
statistical techniques on photometry around compact groups could
applied to many more groups for improved statistics and to look
for systematic trends with group properties.
Furthermore, the spectroscopic data confirm the conclusion of
our photometric analysis that the faint galaxies in our Hickson 
group sample are more diffusely distributed than the brighter galaxies
(Ribeiro et al.\ 1997). The implications of this result are discussed 
in the following section.

	Possible differences between the luminosity functions of 
various types of galaxies are also of interest.
In spectroscopic surveys like ours, a natural division is between
galaxies with and without emission lines. In Figure 3, we plot
the luminosity function for galaxies in our Hickson group sample,
with galaxies in which we detect emission lines now plotted with
different symbols than those in which we do not detect emission
lines. 
This figure shows that the luminosity function for emission-line
galaxies appears to be {\it shallower} than that for galaxies without 
emission lines.

	If confirmed, this result would indicate a difference between
galaxies in Hickson groups and those in the general field, 
as Lin et al.\ (1996b) found that
emission line galaxies have a {\it steeper} faint end slope than galaxies 
without emission lines. Similarly, Loveday et al. (1992) found
that galaxies classified as early-type (less likely to have emission
lines) have shallower faint end slopes than classified as later-type. 
Although this latter result may have been affected by the difficulty 
of classifying galaxies on the available plate material (Marzke et al.\ 1994b),
no previous redshift survey has found that galaxies with emission lines
or of later morphological type, have a shallower faint end slope than 
galaxies without emission lines or of earlier type. A result which might
be similar to ours is that Sandage et al.\ (1985) find that the very
faint end of the luminosity function in Virgo is dominated by dwarf 
ellipticals, which are not known to have emission lines.

	A concern in the comparison of our Hickson group galaxy
luminosity function for emission and non-emission line galaxies
and other surveys is whether the classification of galaxy spectral type
is similar. In our survey, galaxies are classified as emission line
objects if the equivalent width of H$\alpha$ is greater than 6\AA. 
Approximately $60\%$ of the galaxies in our Hickson groups sample
are classified as having emission lines on this basis.
As a comparison, in the red selected LCRS, roughly $50\%$ of the
full sample of galaxies are classified as emission-line objects, 
based on having [OII] equivalent widths of more than 5 \AA. 
In the blue selected samples
of Marzke et al.\ (1994b) and Loveday et al. (1992), about
$70\%$ of the galaxies are classified as late-type on the
basis of morphology. Hickson groups are known to be somewhat
more elliptical-rich than the field (e.g. Hickson, Kindl, \& Huchra\ 1988).
Therefore, the identification of emission-line objects in our 
blue-selected Hickson sample is at least roughly consistent with that 
in other samples. This suggests that the spectral classification itself 
is not responsible for the observed differences in luminosity function 
as a function of spectral type between our Hickson group sample and field 
galaxy samples.

\section{Discussion}

	The primary conclusion of this paper is that the faint end
of the luminosity function of galaxies in Hickson groups is similar
to that found in general field surveys. This result fits well into
the picture that most galaxies in Hickson groups are not significantly
different from those in other environments. There is good evidence
for enhanced merging activity in compact groups (Zepf 1993 and references
therein), as well as peculiarities in the isophotal properties of Hickson 
group ellipticals that may be due to an increased frequency of dynamical 
interactions (e.g.\ Zepf \& Whitmore 1993, Bettoni \& Fasano 1993, 
Mendes de Oliveira \& Hickson 1994, Pildis, Bregman, \& Schombert 1995). 
However, none of these observations indicate that a large fraction of 
the galaxies are strongly affected by their location within Hickson 
groups (Zepf 1995). The absence of evidence for a large fraction of 
ongoing merging might be understood if the Hickson sample is composed 
of groups in a range of dynamical states (Ribeiro et al.\ 1997).

	The slope of the faint end of the luminosity function for
galaxies in our Hickson group sample we find here ($\alpha \simeq -1.0$)
is consistent with our earlier estimate based on galaxy counts in the 
region of the groups and a statistical background correction 
(Ribeiro et al.\ 1994). However, as reviewed by Hickson (1997), 
some other analyses have suggested a depletion of faint galaxies in 
Hickson groups. This work is different than other surveys in two
significant ways. Firstly, our surveys go much deeper, and therefore
provide better leverage on the slope at the faint end.
Secondly, by studying galaxies much fainter than the basis for
Hickson's compact group selection, we avoid many of the potential
biases associated with this selection.

	A result related to this latter point is that we find that
the faint galaxies are more diffusely distributed than the bright galaxies
originally selected by Hickson. Thus, a study of the luminosity function
restricted only to the area on the sky which encloses the bright
galaxies in the group will systematically underestimate the number
of faint galaxies. The effect on the luminosity function can be
significant as we find that the average pairwise radius of the faint
galaxies is about twice that of the bright galaxies.

	The wider spatial distribution of the faint galaxies compared
to the bright galaxies can result from two different effects. One possibility
is the bias inherent in selecting for a compact arrangement of bright
galaxies in the the plane of the sky.  Clearly this favors situations
in which the bright galaxies are aligned to enhance their surface
density. However, since the fainter galaxies are not part of the
selection process, they are not biased in this way. They therefore
may provide a truer representation of the extent of the group.
It is also possible that the more diffuse distribution of
faint galaxies arises from mass segregation. Although there is little
evidence for such an effect in any other system of galaxies, it
is difficult to choose between these two explanations solely on the 
basis of the available data. We note that in either case, 
the true spatial extent of the Hickson groups is underestimated by a factor
of several if only the bright galaxies are studied.

\acknowledgments

We are grateful to Huan Lin for providing the results of the LCRS luminosity 
function study in digital form. This paper benefitted from discussions 
with Ann Zabludoff and from the suggestions of an anonymous referee.
S.E.Z. acknowledges support from NASA through grant
number HF-1055.01-93A awarded by the Space Telescope Science Institute,
which is operated by the Association of Universities for Research in
Astronomy, Inc., for NASA under contract NAS5-26555.
A.L.B. Ribeiro acknowledges the support
of the CAPES. We are grateful for the support of the Cerro Tololo
Inter-American Observatory and STScI at which the spectroscopy and
digitized plate scans were obtained.


\vskip 36pt

\centerline{\bf Figure Captions}

\figcaption[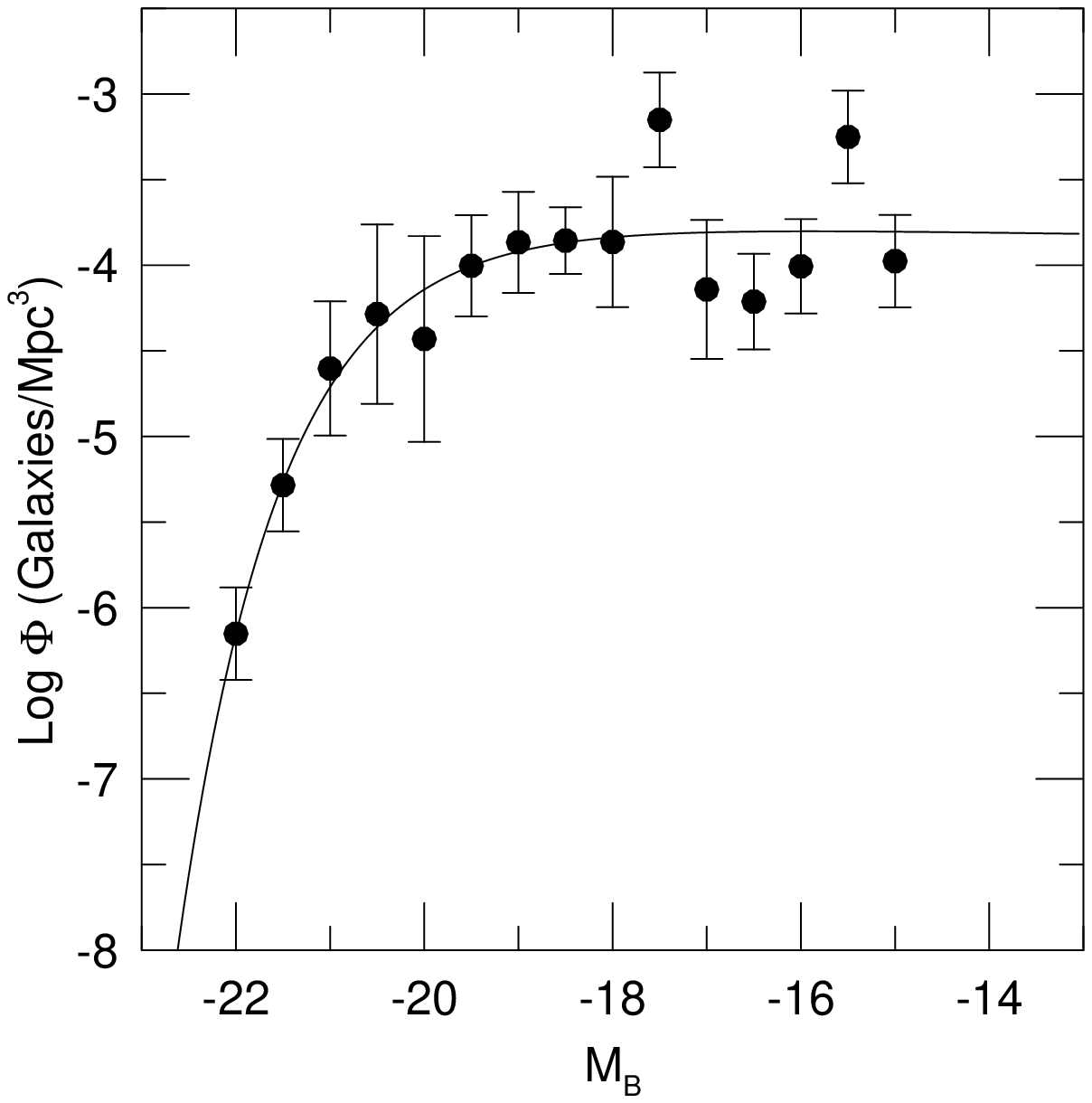]{A plot of the luminosity function of
galaxies in our sample of 17 nearby Hickson groups. The best fitting
Schechter luminosity function is shown as the solid line. 
The data clearly indicate a flat slope for the faint end of the
luminosity function ($\alpha \simeq -1$). We adopt H$_0 = 75~
{\rm km} {\rm s}^{-1} {\rm Mpc}^{-1}$
to determine absolute magnitudes for this and subsequent plots.}
\figcaption[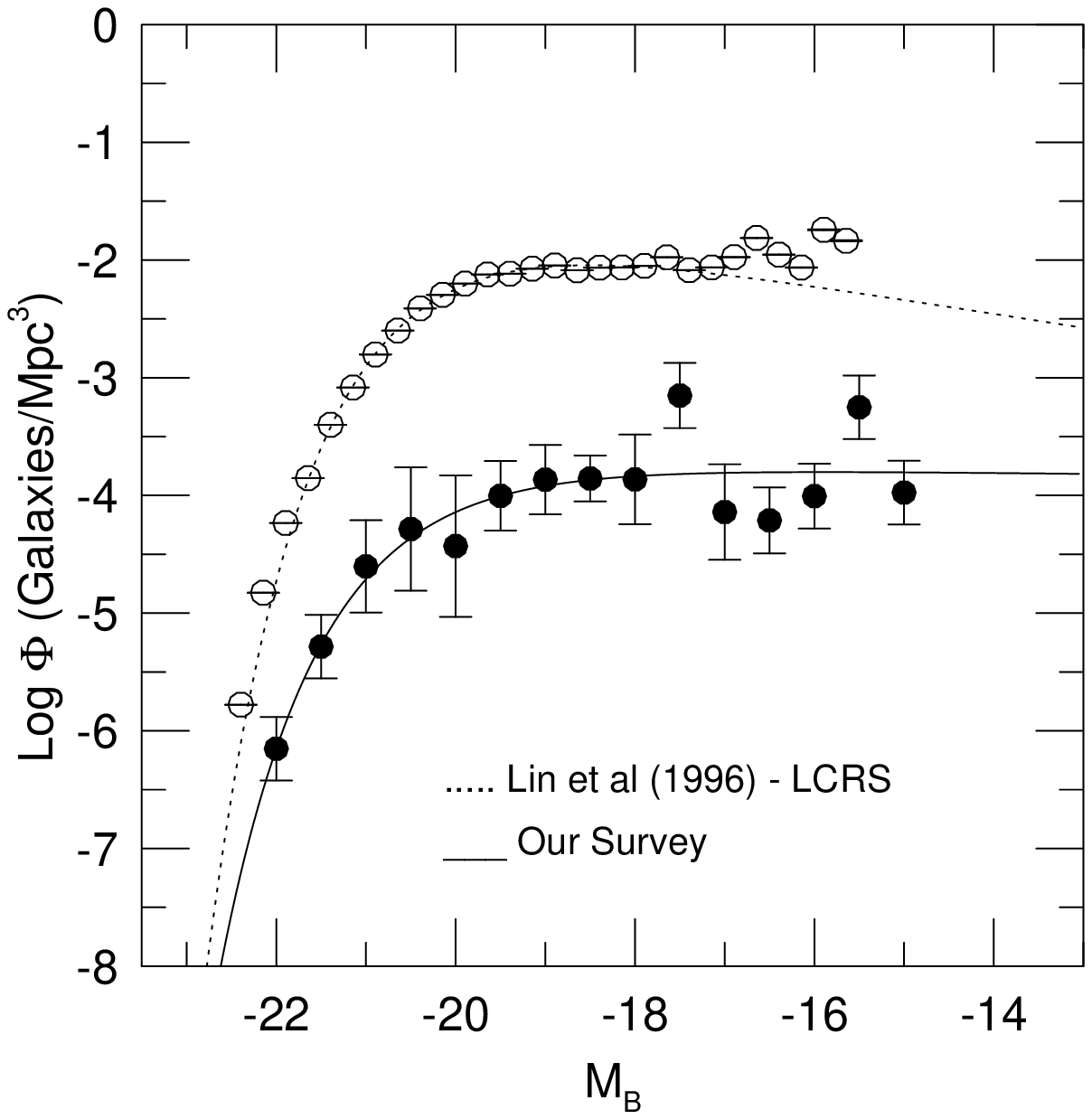]{A comparison of the luminosity function derived
for our sample of galaxies in Hickson groups to that found by Lin
et al. (1995b) for the large Las Campanas redshift survey. The appear
to have a very similar shape. The vertical offset is arbitrary.}
\figcaption[lf_figure3.ps]{A plot of the luminosity function of our
sample of galaxies in Hickson groups, divided by the presence or 
absence of emission lines in the spectrum of the galaxy, where a
galaxy is said to have emission lines if EW (H$\alpha) > 6\AA$. 
This figure shows a decline in the number of emission-line galaxies
at faint magnitudes. Similar surveys of galaxies in other environments
typically find the opposite trend. If confirmed, this result represents
a significant difference between galaxies in compact groups and those 
in other environments.}

\end{document}